\documentclass{jfpc}
\usepackage{url}
\usepackage{listings}
\usepackage[final]{pdfpages}
\usepackage{subfig}
\newcommand{\bugassist}{\textsc{BugAssist}}        
\newcommand{\locF}{\textsc{LocFaults}}        
\annee{2015}
\titre{Un algorithme incrémental dirigé par les flots et basé sur les contraintes pour l'aide à la localisation d'erreurs}

\auteurs{Mohammed Bekkouche \and  Hélène Collavizza \and Michel Rueher} 

\institutions{Univ. Nice Sophia Antipolis, CNRS, I3S, UMR 7271, 06900 Sophia Antipolis, France
}

\mels{\{bekkouche,helen,rueher\}@unice.fr}

\begin{document}
\creationEntete

\begin{resume} 
Dans cet exposé, nous présentons notre algorithme amélioré~\cite{bekkouche15locfaults} de localisation d'erreurs à partir de contre-exemples, \locF{}, basé sur la programmation par contraintes et dirigé par les flots. Cet algorithme analyse les chemins du CFG (Control Flow Graph) du programme erroné pour calculer les sous-ensembles d'instructions suspectes permettant de corriger le programme. En effet, nous générons un système de contraintes pour les chemins du graphe de flot de contrôle pour lesquels au plus $k$ instructions conditionnelles peuvent être erronées. Ensuite, nous calculons les MCS (Minimal Correction Set) de taille limitée sur chacun de ces chemins. La suppression de l'un de ces ensembles de contraintes donne un sous-ensemble satisfiable maximal, en d'autres termes, un sous-ensemble maximal de contraintes satisfaisant la postcondition. Pour calculer les MCS, nous étendons l'algorithme générique proposé par Liffiton et Sakallah  \cite{liffiton2008algorithms,liffiton2013enumerating} dans le but de traiter des programmes avec des instructions numériques plus efficacement. Nous nous intéressons à présenter l'aspect incrémental de ce nouvel algorithme qui n'est pas encore présenté aux JFPC.

Considérons le programme {\tt AbsMinus} (voir fig. \ref{AbsMinus}). Les entrées sont des entiers $\{i, \; j\}$ et la sortie attendue est la valeur absolue de $i-j$. Une erreur a été introduite sur la ligne 10, ainsi pour les données d'entrée $\{i=0, j=1\}$, {\tt AbsMinus} retourne $-1$. La post-condition est juste $result = |i-j|$.

\begin{figure}[!ht]
    \center
\begin{lstlisting}{}
1 class AbsMinus {	
2 /*Il renvoie |i-j|, la valeur absolue de i moins j*/	
3 /*@ ensures 
4  @ ((i < j) ==> (\result == j-i)) && 
5  @ ((i >= j) ==> (\result == i-j)); */
6  int AbsMinus (int i, int j) {
7    int result;
8    int k = 0;
9    if (i <= j) {
10	  k = k+2; }  // erreur: k = k + 2 au lieu de k = k + 1	
11    if (k == 1 && i != j) {
12	  result = j-i; }  
13    else {
14	  result = i-j; }
15    return result;  } }
\end{lstlisting}
\vspace{-0.5cm}
\caption{Le programme AbsMinus}
\label{AbsMinus}
\end{figure}

Le graphe de flot de contrôle (CFG) du programme {\tt AbsMinus} et un chemin erroné sont représentés dans la figure \ref{AbsMinusCFG}. Ce chemin erroné correspondent aux données d'entrée: $\{ i= 0, j=1 \}$.
Tout d'abord, \locF{} collecte sur le chemin \ref{AbsMinusCFG}.(b) l'ensemble de contraintes $ C_1 =\{ i_0=0, j_0=1, k_0=0, k_1=k_0+2,r_1=i_0-j_0 \} $\footnote{Nous utilisons la transformation en forme DSA~\cite{barnett2005weakest} qui assure que chaque variable est affectée une seule fois sur chaque chemin du CFG.}. Puis, \locF{} calcule les MCS de $C_1$. Seulement un MCS peut être trouvé dans $ C_1 $:   $\{ r_1=i_0-j_0 \} $. En d'autres termes, si nous supposons que les instructions conditionnelles sont correctes, la seule instruction suspecte sur ce chemin erroné est l'instruction 14.

\begin{figure}[ht]
\centering
\begin{tabular}{cc}
\subfloat[]{\includegraphics[width=.45\linewidth]{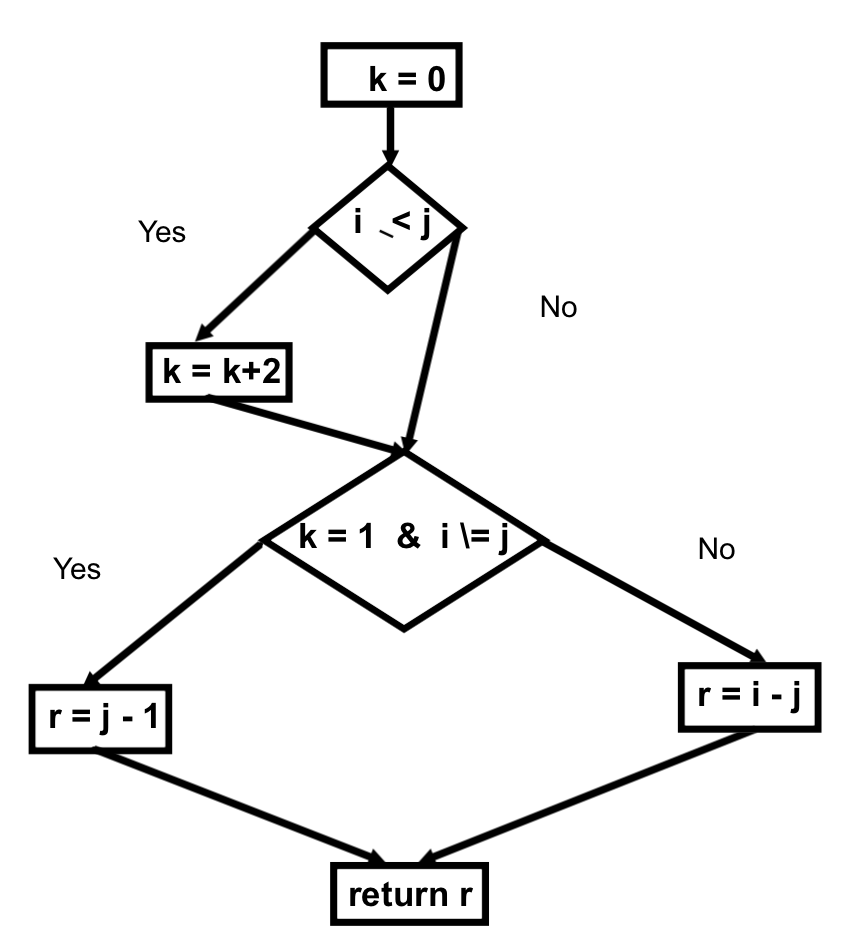} } 
\subfloat[]{\includegraphics[width=.45\linewidth]{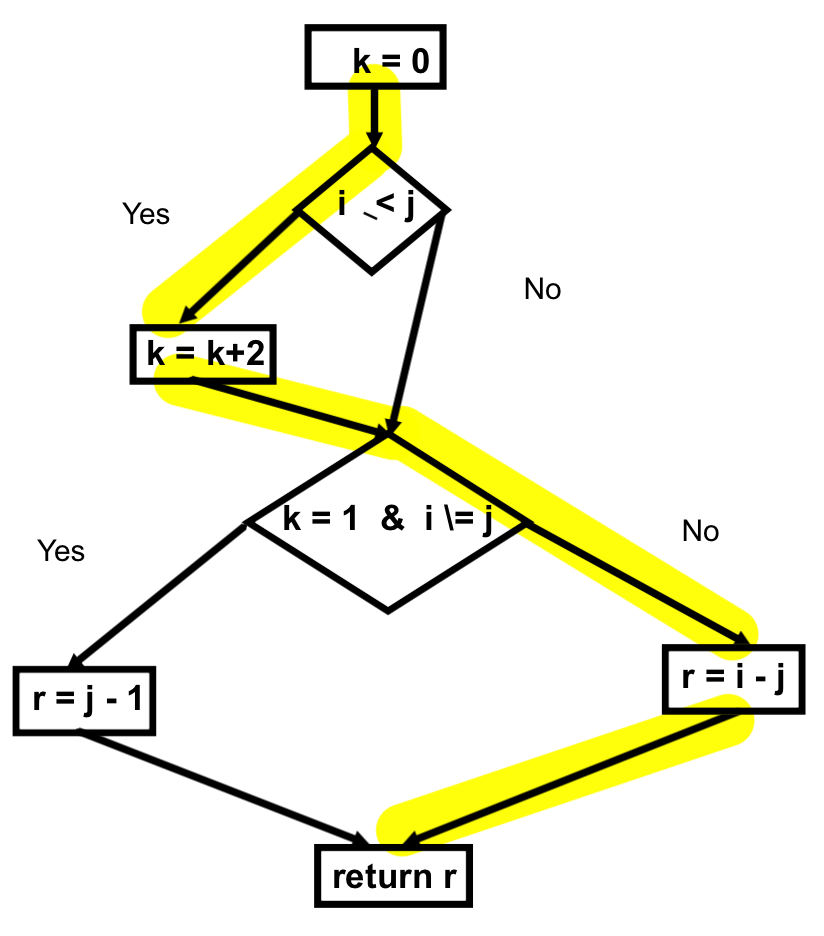}}\\
\end{tabular}
\vspace{-0.2cm}
 \caption{\footnotesize Le CFG et chemin erroné -- Le programme {\tt AbsMinus} 
\label{AbsMinusCFG}}
\end{figure}

Ensuite, \locF{} commence le processus de déviation. La première déviation (voir la figure \ref{AbsMinusD1}.(a), chemin vert) produit encore un chemin qui viole la post-condition, et donc, nous l'ignorons.\
La second déviation (voir la figure \ref{AbsMinusD1}.(b), chemin bleu) produit un chemin qui satisfait la postcondition. Donc, \locF{} collecte les contraintes sur la partie du chemin \ref{AbsMinusD1}.(b)  qui précède la condition déviée, c'est-à-dire $ C_2 =\{ i=0, j=1, k_0=0, k_1=k_0+2 \} $. Puis  \locF{} recherche les MCS de   $ C_2 \cup \neg (k=1 \wedge i \ne j)$; c'est-à-dire nous essayons d'identifier les instructions qui doivent être modifiées afin que le programme aura un chemin satisfaisant la post-condition pour les données d'entrée.
Ainsi, pour cette deuxième déviation deux instructions suspectes sont identifiées:
\begin{itemize}
\item L'instruction conditionnelle sur la ligne 11;
\item L'affectation sur la ligne 10 car la contrainte correspondante est le seul MCS dans $ C_2 \cup \neg (k=1 \wedge i \ne j) $.
\end{itemize}

Puis, \locF{} tente de dévier une seconde condition. Le seul chemin possible est celui où les deux conditions du programme {\tt AbsMinus} sont déviées. Cependant, comme il a le même préfixe que le premier chemin dévié, nous le rejetons.

\vspace{-0.5cm}
\begin{figure}[ht]
\centering
\begin{tabular}{cc}
\subfloat[]{\includegraphics[width=.45\linewidth]{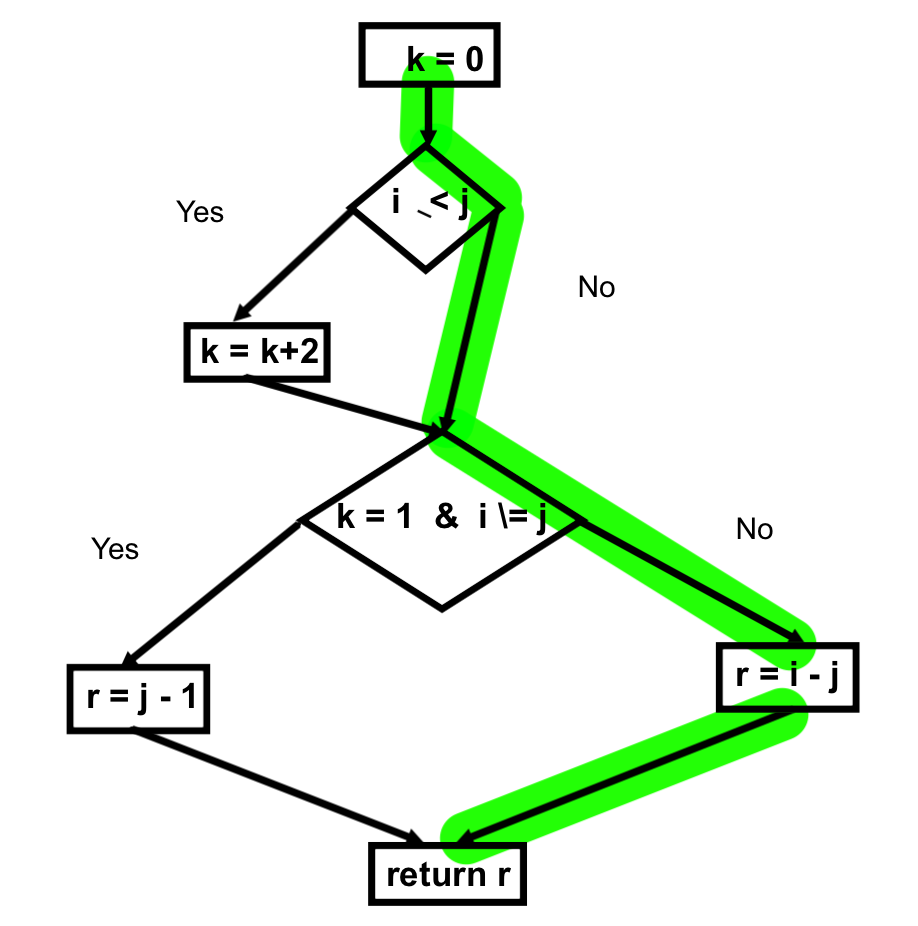} }  
\subfloat[]{\includegraphics[width=.45\linewidth]{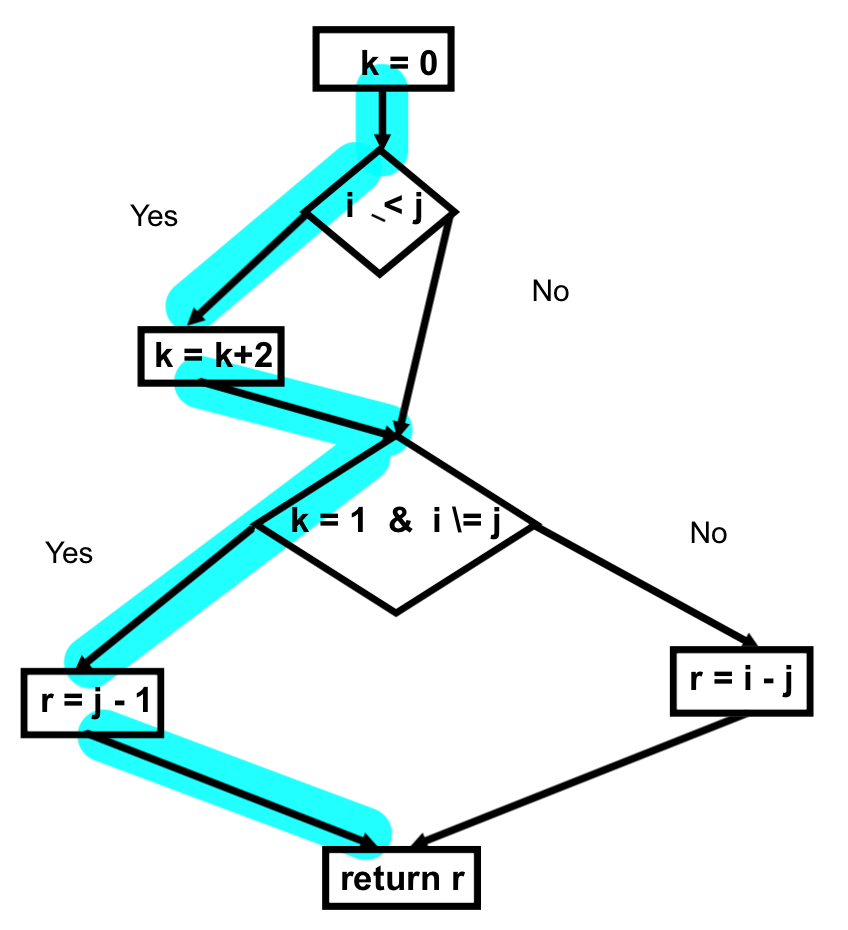}}\\
\end{tabular}
\vspace{-0.2cm}
\caption{\footnotesize Les chemins avec une déviation -- Le programme {\tt AbsMinus}}
\label{AbsMinusD1}
\end{figure}
\vspace{-0.3cm}

Cet exemple montre que \locF{} produit des informations pertinentes et utiles sur chaque chemin erroné. Contrairement à \bugassist{}~\cite{jose2011cause,jose2011bug}, un système de l'état de l'art, il ne fusionne pas toutes les instructions suspectes dans un seul ensemble, ce qui peut être difficile à exploiter par l'utilisateur.

Les entrées de notre algorithme sont le CFG du programme, $CE$: un contre-exemple, $b_{cond}$: une borne sur le nombre de conditions qui sont déviées, et $b_{mcs}$: une borne sur le nombre de MCSs(Minimal Correction Subsets) générés. Grosso modo, notre algorithme explore en profondeur le CFG en utilisant $CE$ pour prendre la branche  \textit{If} ou \textit{Else} de chaque n\oe{}ud conditionnel, et collecte les contraintes qui correspondent aux affectations sur le chemin induit. Il dévie zéro, une ou au plus $b_{cond}$ décisions par rapport au comportement du contre-exemple $CE$. \`A la fin d'un chemin, l'ensemble des contraintes qui ont été collectées est inconsistant, et au plus $b_{mcs}$ MCSs sont calculés sur ce CSP.

Plus précisément, \locF{} procède comme suit:

\begin{itemize}
\item[*] Il propage premièrement $CE$ sur le CFG jusqu'à la fin du chemin initial erroné. Puis, il calcule au plus $b_{mcs}$ MCSs sur le CSP courant. Ce qui représente une première localisation sur le chemin du contre-exemple.

\item[*] Après, \locF{} essaye de dévier une condition. Lorsque le premier n\oe{}ud conditionnel (noté $cond$) est atteint, \locF{} prend la décision opposée à celle induite par $CE$, et continue à propager $CE$ jusqu'au dernier n\oe{}ud dans le CFG. Si le CSP construit à partir du chemin dévié est consistant, il y a deux types d'ensemble d'instructions suspectes:
\begin{itemize}
\item le premier est la condition $cond$ elle-même. En effet, changer la décision pour $cond$ permet à $CE$ de satisfaire la postcondition ;
\item une autre cause possible de l'erreur est une ou plusieurs mauvaises affectations avant $cond$ qui ont produit une décision erronée. Puis, \locF{} calcule aussi au plus $b_{mcs}$ MCSs sur le CSP qui contient les contraintes collectées sur le chemin qui arrivent à $cond$.
\end{itemize}   
Ce processus est répété sur chaque n\oe{}ud conditionnel du chemin du contre-exemple.

\item[*] Un processus similaire est ensuite appliqué pour dévier pour tout $k \leq b_{cond}$ conditions. Pour améliorer l'efficacité, les n\oe{}uds conditionnels qui corrigent le programme sont marqués avec le nombre de déviations qui ont été faites avant d'avoir été atteints. Pour une étape donnée $k$, si le changement de la décision d'un n\oe{}ud conditionnel $cond$ marqué avec la valeur $k'$ avec $k' \leq k$ corrige le programme, cette condition est ignorée. En d'autres termes, nous considérons seulement la première fois où un n\oe{}ud conditionnel corrige le programme.
\end{itemize}

Cet algorithme incrémental basé sur les flots est un bon moyen pour aider le programmeur à la chasse aux bugs car il localise les erreurs autour du chemin du contre-exemple. Nos résultats ont confirmé que les temps de cet algorithme sont meilleurs par rapport à ceux qui correspondent à l'algorithme que nous avons présenté aux JFPC 2014 dans~\cite{bekkouche14locfaults} \footnote{Les résultats qui présentent les temps de calcul des deux versions de \locF{}, non-incrémentale et incrémentale, pour des programmes sans boucles sont disponibles à l'adresse \url{http://www.i3s.unice.fr/~bekkouch/Bench_Mohammed.html#rsba}}. Les sous-ensembles d'instructions suspectes fournis sont plus pertinents pour l'utilisateur. Dans le cadre de notre travaux futurs, nous envisageons de confirmer nos résultats sur des programmes avec plusieurs boucles complexes (voir nos premiers résultats dans~\cite{bekkouche2015exploration}). Nous envisageons de comparer les performances de \locF{} avec des méthodes statistiques existantes ; par exemple: Tarantula~\cite{jones2005empirical,jones2002visualization}, Ochiai~\cite{abreu2007accuracy}, AMPLE~\cite{abreu2007accuracy}, Pinpoint~\cite{chen2002pinpoint}. Nous développons une version interactive de notre outil qui fournit les sous-ensembles suspects l'un après l'autre : nous voulons tirer profit des connaissances de l'utilisateur pour sélectionner les conditions qui doivent être déviées. Nous réfléchissons sur comment étendre notre méthode pour supporter les instructions numériques avec calcul sur les flottants. 

\end{resume}



\end{document}